\documentclass[12pt]{article}
\usepackage{amsmath,amssymb,epsfig}

\def\K{{\mathcal K}}
\def\G{{\mathcal G}}
\def\be{\begin{equation}}
\def\ee{\end{equation}}
\def\lp{\ell_P}
\def\R{{\mathcal R}}
\def\R{{\mathcal R}}

\def\L{{\mathcal L}}
\def\H{{\mathcal H}}
\def\L{{\mathcal L}}
\def\K{{\mathcal K}}
\def\be{\begin{equation}}
\def\ee{\end{equation}}
\def\lp{\ell_P}

\def\a{\alpha}
\def\b{\beta}
\def\g{\gamma}
\def\d{\delta}

\def\r{\rho}

\def\m{\mu}\def\n{\nu}
\def\bnabla{\bar\nabla}

\def\bbox{\overline{\Box}}\def\beq{\begin{eqnarray}}\def\eeq{\end{eqnarray}}

\begin{document}
\title{On c-theorems in arbitrary dimensions}

\author{{ Arpan Bhattacharyya}$^{1}$,
{ Ling-Yan Hung}$^{2}$, {Kallol Sen}$^{1}$ and {Aninda Sinha}$^{1}$\\
\it $^1$Centre for High Energy Physics, Indian Institute of Science,  Bangalore, India.\\
\it $^2$Perimeter Institute for Theoretical Physics, Waterloo, Canada.}
\maketitle
\begin{abstract}
The dilaton action in  3+1 dimensions plays a crucial role in the proof of the a-theorem. This action arises using Wess-Zumino consistency conditions and crucially relies on the existence of the trace anomaly. Since there are no anomalies in odd dimensions, it is interesting to ask how such an action could arise otherwise. Motivated by this we use the AdS/CFT correspondence to examine both even and odd dimensional CFTs.  We find that in even dimensions, by promoting the cut-off to a field, one can get an action for this field which coincides with the WZ action in flat space. In three dimensions, we observe that by finding an exact Hamilton-Jacobi counterterm, one can find a non-polynomial action which is invariant under global Weyl rescalings. We comment on how this finding is tied up with the F-theorem conjectures.
\end{abstract}
\tableofcontents

\section{Introduction}
Zamolodchikov's c-theorem \cite{zamo} in the context of 1+1 dimensional unitary conformal field theories is a profound and elegant result. It states that in the space of couplings, there is a quantity called the central charge that monotonically decreases along an RG flow. This is a realization of the intuition that when one integrates out the high energy modes in a Wilsonian sense, the number of degrees of freedom should decrease. The question automatically arises as to how this can generalize to other dimensions--see \cite{early} for earlier work on this. On the holographic side, this has been studied originally in \cite{cthor} and in recent times for example in \cite{myersme, oddrefs,others}. Last year there was an elegant proof of the so-called ``a-theorem" in the context of 3+1 dimensional unitary conformal field theories by Komargodski and Schwimmer \cite{Komargodski:2011vj}.

One key element in this proof was to write down the action for the Nambu-Goldstone boson corresponding to a spontaneously broken conformal symmetry. When conformal symmetry is broken however, as explained in \cite{Komargodski:2011xv, Luty:2012ww} one could equivalently place the theory on a conformally flat metric $e^{-2\sigma}\eta$, such that $\sigma$
plays the role of a dilaton, and the effective action $W[\sigma]$ of which emerges as we integrate out the quantum fields.
In this paper, we will ask the question: is there a natural way to incorporate these ideas  in AdS/CFT, such that we can construct
and follow the change of $W[\sigma]$ as we move along the renormalization group flow holographically?

By this we mean specifically the following. On the gravity side, the usual starting point is the Einstein-Hilbert action together with the Gibbons-Hawking term and counterterms which are needed to cancel UV divergences\footnote{One could consider adding additional matter fields such as scalar fields as well, but around AdS spaces, these fields are frozen.}.
\be\label{totact}
I_{tot}=I_{EH}+I_{GH}+I_{ct}\,,
\ee
The counterterm action is made up of the curvature invariants of the boundary metric. If there is any hope of extending this calculation for odd-dimensional CFTs, one should not use the Wess-Zumino consistency conditions which crucially depend on the existence of an anomaly.
Using eq.(\ref{totact}) as the starting point, how do we get $W[\sigma]$ ?
The answer that we find is the following. Consider without loss of generality
the case where the boundary space is flat and we are working with
the Poincar\'e metric. The radial cut-off, usually taken near the AdS boundary,
in a sense plays the role of an RG scale \cite{de Boer:1999xf, Heemskerk:2010hk, Faulkner:2010jy}.
If we promote the radial coordinate of the cut-off surface
to a field\footnote{Sometimes called the radion.} (in a manner to be made specific) and do a derivative expansion in this field, then
from $I_{tot}$ we will recover the dilaton action around flat space.
In particular, since we will consider pure AdS in this paper and if the cut-off surface stays close to
the AdS boundary, the only source of conformal symmetry breaking comes from conformal anomaly in even
dimensions. Not surprisingly therefore, in even dimensions the dependence on $\sigma$ comes solely from the WZ action that
encodes the anomaly \cite{st, Komargodski:2011vj}. Note that if the theory is $d$ dimensional,
then the Wess-Zumino (WZ) action for the dilaton starts at $d$-derivative order.
The fact that the WZ action can be recovered by changing the cut-off surface
has been noticed before in the case of Einstein gravity in AdS$_3$/CFT$_2$ \cite{Krasnov:2001cu,Hung:2011nu}.
We clarify and generalize it systematically to higher dimensions and different gravity theories. We point out
several key features and observations in our calculation:
\begin{itemize}
\item  Only surface terms defined at the cut-off, namely the Gibbons-Hawking term,
and the counterterms comprising of curvature invariants that get rid of power law divergences
contribute to $W[\sigma]$. For the anomaly, no local counterterm made of curvature invariants can remove it. Even if we added an explicit term like $\int \log \epsilon \langle T_a^a \rangle $ by hand \cite{de Haro:2000xn} (where $\epsilon$ is a UV cut-off), this would not affect our result.
\item In order to distinguish between the Euler anomaly coefficient from the others, we have considered Gauss-Bonnet gravity duals for 3+1 and 5+1 dimensional CFTs \cite{gbholo}. For these theories, the generalized Gibbons-Hawking term is known \cite{Liu:2008zf} which facilitates our calculation. In 1+1 dimensions,  we have also considered New Massive Gravity \cite{nmg1} which does not lead to 2 derivative equations of motion and for which again the Gibbons-Hawking term is known \cite{hohm}. Even for this theory, our finding is consistent with what is expected from the dilaton action.
\end{itemize}

How can we generalize these to odd dimensions? Motivated by this question, we consider counterterms in AdS$_4$/CFT$_3$ in Einstein gravity. The systematic way to derive counterterms on the gravity side is to solve the Hamiltonian constraint \cite{kraus}. A recursive way of finding the counterterm action can be derived. In even dimensions, the existence of the anomaly leads to a breakdown in the recursion relations. However, in odd dimensions, one can in principle resum the entire series. We observe that the DBI type counterterm in the CFT$_3$ context that was proposed in \cite{js} is an exact solution to the Hamiltonian constraint if we consider $S^3$ or $S^1\times S^2$ as the Euclidean boundary (also their analytic continuations). We show that the leading deviation from conformality can be written in terms of the square of the Cotton tensor which is known to play the role of the Weyl tensor in 3 dimensions. We show that when one considers this action around the origin of AdS$_4$ with $S^3$ slicing, %JH%
 then the resulting non-polynomial action is invariant under global Weyl transformations. This points at a heretofore unobserved universality related to the F-theorem conjectures \cite{oddrefs}. We study various properties of this counterterm action. This may be a useful starting point to generalize the Komargodski-Schwimmer proof to odd dimensions.

This paper is organized as follows. In section 2, we compute the dilaton action for Gauss-Bonnet gravity duals for even dimensional CFTs. We turn to the odd dimensional cases in section 3, focusing on 3 dimensions. In appendix A, we review recursion relations for the counterterm action which follow from the Hamiltonian constraint. In appendix B, we show how the pole terms in even dimensions can be used to extract the WZ action. We will denote the boundary dimensions by $d$. We will further denote the length scale appearing in the cosmological constant by $L$ and the AdS radius by $\tilde L$ (which is the same as $L$ in Einstein gravity).
\section{Dilaton actions in GB holography--Even dimensions}
 In $d+1$-dimensional spacetimes the full Euclidean action has three parts,
\be
I_{tot}=I_{bulk}+I_{surf}+I_{ct}\\.
\ee
For Gauss-Bonnet gravity (we will follow the conventions in \cite{gbholo}),
\be
I_{bulk}=-\frac{1}{2 \lp^{d-1}}\int d^{d+1}x\sqrt{g}(R+\frac{d(d-1)}{L^2}+
 \frac{L^{2}\lambda}{(d-2)(d-3)} GB)\,,\\
\ee
where, $GB=R_{\mu\nu\rho\sigma}R^{\mu\nu\rho\sigma}-4R_{\mu\nu}R^{\mu\nu}+R^{2}$.
\be
I_{surf}=I_{GH}^{(1)}+I_{GH}^{(2)}\,,\\
\ee
where,
\be
I_{GH}^{(1)}=-\frac{1}{\lp^{d-1}}\int d^{d}x\sqrt{h} \K\,,\\
\ee
and
\be
I_{GH}^{(2)}=\frac{2}{\lp^{d-1}}\frac{\lambda L^2}{(d-2)(d-3)}\int d^{d}x\sqrt{h}[2\G_{ab}\K^{ab}+\frac{1}{3}(\K^{3}-3\K\K_{ab}\K^{ab}+2\K_{a}^{b}\K_{b}^{c}\K^{a}_{c})]\,.\\
\ee
Here  $\K_{ab}= h^{c}_{a}h^{d}_{b}\nabla_{c}\hat n_{d}$  is the extrinsic curvature defined on the boundary with the induced metric $h_{ab}=g_{ab}-\hat n_{a}\hat n_{b}$ , where $\hat n_{a}$ is the unit normal  and $\K= h^{ab}\nabla_{a}\hat n_{b}$ being the trace of the extrinsic curvature.
\be\label{ict}
I_{ct}=\frac{1}{\lp^{d-1}}\int d^{d}x \sqrt{h} [c_{1}\frac{d-1}{\tilde L} +c_{2} \frac{\tilde L}{2(d-2)} \R+c_{3}\frac{\tilde L^{3}}{2(d-4)(d-2)^{2}}(\R_{ab}\R^{ab}-\frac{d}{4(d-1)}\R^{2})+.......]\,,\\
\ee
is the generalization of the  counterterm action in Einstein gravity \cite{Emparan:1999pm}
where, $\frac{L^{2}}{\tilde L^{2}}=f_{\infty}$  with $\tilde L$ being the AdS radius and the equations of motion giving $1-f_{\infty}+f_{\infty}^{2}\lambda=0$. One has to consider for $d=2$  only the first term, $d=4$ the first two terms and $d=6$ all three terms. The existence of simple poles at $d=2$ and $d=4$ points at the existence of the conformal anomaly.
The coefficients in the counterterm action are:
\begin{align}
\begin{split}
 c_{1}= 1- \frac{2}{3}f_{\infty}\lambda\,,\quad c_{2}=1+2 f_{\infty}\lambda\,,\quad c_{3}=1-6f_{\infty}\lambda\,.\\
 \end{split}
\end{align}
 Curiously the coefficient $c_3$ is proportional to the  Euler anomaly for $d=4$.  This appears to be some kind of dimensional regularization since there is a simple pole at $d=4$ in the counterterm action (see also \cite{Imbimbo:1999bj}). Furthermore, the $c_1$ coefficient is again related to the Euler anomaly in New Massive Gravity which follows from the general formula for  the central charge \cite{myersme, Imbimbo:1999bj}
\begin{equation}\label{astar}
a_{d}^{*}=\frac{\pi^{d/2} L^{d-1}}{\Gamma(d/2)\lp^{d-1}f_{\infty}^{(d-1)/2}}(1-\frac{2(d-1)}{d-3} f_{\infty}\lambda)\,.
\end{equation}
\par
We will later on see that this anomaly coefficient formula applies to the new massive gravity theory which was argued to share certain features with the Gauss-Bonnet theory in \cite{sinha1}.

For $d+1$ dimensional AdS spaces with Euclidean signature, the metric in Poincar\'e  coordinates is
\be
ds^{2}=\tilde L^{2}\frac{dz^{2}+\sum_{i=1}^{d}{dx^{i}}^{2}}{z^{2}}\,.\\
\ee
For this $R=-\frac{d(d+1)}{\tilde L^{2}}$ and we will take the cut-off as $$ z=e^{\sigma(x^{i})}, $$ where $i=1,..,d$. We will now consider various dimensionalities one by one below. Before we begin, we wish to emphasise that we {\it do not} need to use the simple pole terms in eq.(\ref{ict}) in our construction unlike the method in \cite{Imbimbo:1999bj}.
\subsection{d=2}
For $D=3$ there is no Gauss-Bonnet term as it vanishes identically. We will consider the case of New Massive Gravity in the next section.
The total action after a derivative expansion takes the form
\be
I_{tot}=- \frac{1}{2\lp}\int d^{2}x((\partial \sigma)^{2}+2\partial^{2} \sigma)\epsilon^{2}+\cdots \,,\\
\ee
where, $\epsilon$ is introduced to keep track of the order of the derivative. Other higher order terms denoted by $\cdots$ vanish in the $ z\rightarrow \infty$ limit. The laplacian term is a surface term and can be dropped.
%\be
%\int d^{2}x \partial^{2}\sigma=\int d^{2}x \partial(\partial \sigma) \,.\\
%\ee
Finally we get,
\be
I_{tot}=\frac{a^{*}_{1}}{\pi }\int d^{2}x \mathcal{L}\,,
\ee
where
\begin{equation}
\mathcal{L}=-\frac{1}{2} (\partial \sigma)^{2}\,.
\end{equation}
\subsection{d=4}
In this case, the total action with the GB term is in a derivative expansion by
\be
I_{tot}=\frac{L^{3}}{2 \lp^{3} f_{\infty}^{3/2}}(1-6f_{\infty}\lambda)\int d^{4}x(\partial(e^{-2\sigma}\partial \sigma) \epsilon^{2}+\frac{L^{3}}{\lp^{3}f_{\infty}^{3/2}}\int d^{4}x( \sum_{i=1}^{5}a_{i} t_{i })\epsilon^{4}+\cdots\,,\\
\ee
where, $a_{i}, t_{i}$ 's are listed below:
\begin{align}
\begin{split}
a_{1}&=\frac{-1+14 f_{\infty} \lambda}{4}\,,\quad a_{2}=-1+10 f_{\infty}\lambda\,,\quad a_{3}= \frac{-1+ 14 f_{\infty}\lambda}{4}\,,\\
a_{4}&=-\frac{1-6f_{\infty}\lambda}{8}\,,\quad a_{5}=\frac{1-14 f_{\infty} \lambda}{4}\,,\\
\end{split}
\end{align}
and
\begin{align}
\begin{split}
t_{1}&=(\partial_a \partial_b \sigma)^{2}\,,\quad
t_{2}=\partial_{a}\partial_{b}\sigma\partial^{a}\sigma\partial^{b}\sigma\,,\quad
t_{3}=\partial^{2}\sigma(\partial \sigma)^{2}\,,\quad
t_{4}=( \partial \sigma)^{4}\,,\quad
t_{5}=(\partial^{2}\sigma)^{2}\,.\\
\end{split}
\end{align}
After integration by parts, up to a surface term,
\begin{align}
\begin{split}
t_{1}=t_{5}\,,\quad
t_{2}=-\frac{1}{2}t_{3}\,.\\
\end{split}
\end{align}
Using these and throwing away all the surface terms,
\be
I_{tot}=\frac{a^{*}_{4}}{\pi^{2}}\int d^{4}x \mathcal{L}\,,\\
\ee
where,
\be
\mathcal{L}=-\frac{1}{8} (\partial \sigma )^{4}+\frac{1}{4} \partial^{2} \sigma(\partial \sigma)^{2}\,.\\
\ee
This is precisely the WZ action for the dilaton in flat space \cite{st, Komargodski:2011vj}. Here we see from AdS/CFT that the coefficient of the dilaton action is precisely the Euler anomaly.
Let us comment on the limit $\lambda \to 0, f_\infty \to 1$, where the theory reduces to Einstein gravity.
In that case, the dual CFT is $\mathcal{N}=4, SU(N)$ SYM theory, and we have
\be
a_4^* = \frac{\pi^{2} L^{3}}{\Gamma(2)\lp^{3}} = \frac{N^2}{4},
\ee
where we have made use of the dictionary
\be
\frac{L^4}{{\alpha'}^2} = 4\pi g_s N\,, \qquad 2\lp^3 = 8\pi G_5 = \frac{16\pi^3 g_s^2 {\alpha'}^4}{L^5}\,.
\ee
Here $\alpha'$ is the string length squared. The WZ action thus recovers the well known result for the
$a$-anomaly of $\mathcal{N}=4, SU(N)$ SYM theory.
\subsection{d=6}
The total action with the GB term in a derivative expansion is given by
\begin{align}
\begin{split}
I_{tot}&=\frac{L^{5}}{4\lp^{5} f_{\infty}^{5/2}}(1-\frac{10}{3}f_{\infty}\lambda)\int d^{6}x \partial(e^{-4\sigma}\partial \sigma)\epsilon^{2}+\frac{L^{5}}{8\lp^{5} f_{\infty}^{5/2}}(1-\frac{10}{3}f_{\infty}\lambda)\int d^{6} x\partial_{a}(e^{-2\sigma}\partial_{b}\sigma\partial^{b}\sigma\partial^{a}\sigma)\epsilon^{4}\\
&+\frac{L^{5}}{\lp^{5}f_{\infty}^{5/2}}\int d^{6}x( \sum_{i=1}^{6}a_{i} t_{i })\epsilon^{6}+\cdots\,.\\
\end{split}
\end{align}
where, $a_{i}, t_{i}$ 's are listed below .
\begin{align}
\begin{split}
a_{1}&= \frac{9- 38 f_{\infty}\lambda }{48}\,,\quad a_{2}=\frac{-9+38 f_{\infty}\lambda}{72}\,,\quad a_{3}=\frac{-15+58 f_{\infty}\lambda}{24}\\
a_{4}&=\frac{-9+38 f_{\infty}\lambda}{144}\,,\quad a_{5}=\frac{1-4 f_{\infty}\lambda}{4}\,,\quad a_{6}=\frac{-1+4 f_{\infty}\lambda}{4}\\
a_{7}&=\frac{-27+106 f_{\infty}\lambda}{96}\,,\quad a_{8}=\frac{15-58 f_{\infty}\lambda}{24}\,,\quad a_{9}=\frac{-9+34 f_{\infty}\lambda}{6}\\
a_{10}&=\frac{-1+\frac{10}{3} f_{\infty}\lambda}{16}\,,\\
\end{split}
\end{align}
and
\begin{align} \label{t6}
\begin{split}
t_{1}&=\partial ^2\sigma \left(\partial _a\partial _b\sigma\right){}^2\,,\quad
t_{2}=\partial _a\partial ^b\sigma\partial _b\partial ^c\sigma\partial _c\partial ^a\sigma\,,\quad
t_{3}=\left(\partial _a\partial _b\sigma\right)\left(\partial ^a\partial _c\sigma\right) \partial ^c\sigma \partial ^b\sigma\,,\\
t_{4}&=\left(\partial ^2\sigma\right)^3\,,\quad
t_{5}=\left(\partial ^2\sigma\right)^2(\partial \sigma)^2\,,\quad
t_{6}=\left(\partial _a\partial _b\sigma\right){}^2(\partial \sigma)^2\,,\quad
t_{7}=(\partial \sigma)^4\partial ^2\sigma\,,\\
t_{8}&=\partial ^2\sigma\partial _a\partial _b\sigma \partial ^a\sigma \partial ^b\sigma\,,\quad
t_{9}=(\partial \sigma)^2\partial _a\partial _b\sigma \partial ^a\sigma \partial ^b\sigma\,,\quad
t_{10}=(\partial \sigma)^6\,.\\
\end{split}
\end{align}
After integration by parts, upto a surface term,
\begin{align}\label{t6der}
\begin{split}
t_{2}&=\frac{3}{2}t_{1}-\frac{1}{2}t_{4}\,,\quad
t_{3}=t_{8}+\frac{1}{2}t_{5}-\frac{1}{2}t_{6}\,,\quad
t_{9}=-\frac{1}{4} t_{7}\,.\\
\end{split}
\end{align}
Using these and throwing away all the surface terms we get
\be
I_{tot}=\frac{a^{*}_{6}}{\pi^{3}}\int d^{6}x\mathcal{L}\,,\\
\ee
where,
\be
\mathcal{L}= -\frac{1}{8} (\partial \sigma)^6- \frac{1}{8} \left(\left(\partial ^2\sigma\right)^2(\partial \sigma)^2\right) +\frac{3}{16} (\partial \sigma)^4\partial ^2\sigma +\frac{1}{8} \left(\partial _a\partial _b\sigma\right){}^2(\partial \sigma)^2 \,.\\
\ee
This is precisely the WZ action for the dilaton in 6d-flat space \cite{Elvang:2012st}.
Here, we again consider the limit of Einstein gravity,
whose CFT dual corresponds to the $6d$ (2,0) $SU(N)$ theory on
an M$_5$ brane.
In this case,
\be
a_6^* = \frac{\pi^{3} L^{5}}{\Gamma(3)\lp^{5}} = \frac{N^3}{3}.
\ee

Here, we have made use of the following relations
\be
L= 2(\pi N)^{\frac{1}{3}} {\lp}_{d=11}\,, \qquad \lp^5 = \frac{48 \pi^6({\lp}_{d=11})^{5}}{(\pi N)^{4/3} }.
\ee
Again, we recover the famous leading $N$ result of the $a$-anomaly of the (2,0) theory in the large $N$ limit.

\subsection{Dilaton action for NMG in $D=3$}
In the previous section, we considered GB holography where the bulk equations of motion were two derivative. Since the GB term vanished in 2+1 bulk dimensions, the question naturally arises if one can go beyond Einstein gravity in this case. Fortunately the answer is yes. The main obstacle in going beyond Einstein gravity in our approach is the knowledge of the generalized Gibbons-Hawking term. In the case of New Massive Gravity, the generalized Gibbons-Hawking term is in fact known \cite{hohm}. Furthermore, the bulk equations of motion are not two derivative which makes this case all the more interesting.
In this case\footnote{Following the conventions in \cite{sinha1} which are consistent with $1-f_\infty+\lambda f_\infty^2=0$.},
\be
I_{bulk}=-\frac{1}{2\lp}\int d^{3}x \sqrt{g}(R+\frac{2}{L^{2}}+4\lambda L^{2} (R_{\mu \nu}R^{\mu\nu}-\frac{3}{8}R^{2})\,,\\
\ee
where, $R=-\frac{6}{\tilde L^{2}}$ for $AdS_{3}$\,.
\be
I_{surf}=\frac{1}{2\lp}\int d^{2}x\sqrt{h}(-2 \K-\hat f^{ab}\K_{ab}+\hat f \K)\,.\\
\ee
This the generalized Gibbons-Hawking term. We explain the notation below \cite{hohm}.
For a general metric after an ADM decomposition\,, we can write it as\,,
\be
ds^{2}=N dr^{2}+h_{ab}(dx^{a}+N^{a}dr)(dx^{b}+N^{b}dr)\,.\\
\ee
Here, $a,b$ are the boundary indices. Similarly after a 2+1 split\,,$$f_{\mu \nu}=8\lambda L^{2}(R_{\mu\nu}-\frac{1}{4}R g_{\mu\nu})$$ can be written in the following form\,,
\be
f^{\mu\nu}=\left(
\begin{array}{cc}
 f^{\text{ab}} & h^a \\
 h^b & s
\end{array}
\right)
\ee
Then,
\be
\hat f^{ab}= f^{ab}+2h^{(a} N^{b)}+sN^{a}N^{b}\,,\\
\ee
and $$\hat f =h_{ab}f^{ab}$$
For $AdS_{3}$, $$N^{a}=0\quad h^{a}=0.$$ Also  $f_{\mu\nu}$ is a  $3\times 3$ metric of the form,
\be
f_{\mu\nu}=\left(
\begin{array}{ccc}
 f_{\text{rr}} & 0 & 0 \\
 0 & f_{\text{tt}} & 0 \\
 0 & 0 & f_{\text{xx}}
\end{array}
\right)
\ee
In this case $$\hat f^{ab}=f^{ab}.$$ So  we can obtain $f_{ab} $ from $f_{\mu\nu}$ as,
\be
f_{ab}=f_{\mu\nu}e^{\mu}_{a}e^{\nu}_{b}\,,\\
\ee
where, $e^{\mu}_{a}= \frac{\partial x^{\mu}}{\partial x^{a}}$. The counterterm action works out to be
\be
I_{ct}=\frac{1}{\lp} \int d^{2}x \sqrt{h}c_{1} \frac{D-2}{\tilde L}\,.\\
\ee
where, $c_{1}=1+2\lambda f_{\infty}$. So full action,
\be
I_{tot}=I_{bulk}+I_{surf}+I_{ct}\,.\\
\ee
In this case also $\frac{L^{2}}{\tilde L^{2}}=f_{\infty}$ and $ 1-f_{\infty}+f_{\infty}\lambda^{2}=0$. Proceeding as before\,, after a derivative expansion and throwing  away the surface
term,
\be
I_{tot}=\frac{a^{*}_{2}}{\pi^{2}}\int d^{2}x\left(-\frac{1}{2}(\partial \sigma)^{2}\right)\,,\\
\ee
where $a^*_2$ follows from eq.(\ref{astar}). This is also the Euler anomaly as was shown in \cite{sinha1}. It is gratifying to note that even in this case the dilaton action works out correctly. It will be interesting to see if this generalization can be shown to persist in the quasi-topological gravity theory \cite{quasitop} and the DBI extension of NMG \cite{tekinnmg}.
%\begin{align}
%\begin{split}
%\bf{a^{*}_{D-1}} &=\pi^{2}\frac{L^{D-2}}{\lp^{D-2} f^{(D-2)/2}_{\infty}}(1-2\frac{D-2}{D-4} \lambda f_{\infty})\,,\\
%\end{split}
%\end{align}
%and $$\mathcal{L}= -\frac{1}{2}(\partial \sigma)^{2}.$$

\section{Odd dimensional CFTs--The $d=3$ case}
While our results for even-$d$ may have been anticipated from \cite{Imbimbo:1999bj} and in a sense serves as a check that we understand the building blocks in AdS/CFT, the odd-$d$ case is murkier and hence more interesting. Is there an analog of the WZ action in odd dimensions? There are no anomalies in odd dimensional CFTs so at first sight the answer appears to be no. Let us step back and ask a somewhat simpler question. Consider $d=4$. The existence of the trace anomaly leads to pole terms  in the counterterm action eq.(\ref{ict}) namely
\be\label{ipole}
I_{pole}=c_{3}\frac{\tilde L^{3}}{2(d-4)(d-2)^{2}}\frac{1}{\lp^{3}}\int d^{4}x \sqrt{h} [(\R_{ab}\R^{ab}-\frac{d}{4(d-1)}\R^{2})]\,.\\
\ee
The integrand is invariant under global Weyl transformations $\tilde h_{ab}=e^{2\omega} h_{ab}$. In general any term $\sqrt{h} \R^n$ transforms with a factor of $e^{(d-2n)\omega}$. This fact demonstrates that {\it a local} gravitational term can only be global Weyl invariant in {\it even} dimensions. Moreover, due to the existence of the simple pole at $d=4$, $I_{pole}$ is in fact {\it not} invariant under global Weyl transformation although the integrand is. The pole factor leads to a term that is linear in $\omega$ upon variation which can be seen by expanding the $e^{(d-4)\omega}$. This is the origin of the trace anomaly in even dimensions--we review the $d=4$ calculation in appendix B.

Now let us consider $d=3$. Firstly, we do not expect any trace anomaly in odd dimensions. So the next question that we can ask is if there is any piece in the counterterm action that is invariant under global Weyl transformation. Since a local polynomial curvature invariant cannot be invariant under global Weyl transformation we turn to the simplest generalization--a non-polynomial object. If we consider terms like $\sqrt{h \R^3}$, these will be invariant under global Weyl transformation. Now the question arises why should we consider such non-polynomial objects? Firstly we should emphasise, that such an action cannot arise by expanding around the `usual' boundary ($z\rightarrow 0$) of planar AdS which will necessarily give rise to a local action in odd dimensions. Rather we will find that expanding around the centre of Euclidean AdS with $S^3$ boundary appears to be a natural way to get this non-polynomial action. In the bulk calculation of the Euclidean partition function, this is precisely the way to get the finite part which has been conjectured to play the role of the $c$-central charge with monotonicity properties \cite{oddrefs}. As we will argue next, such non-polynomial counterterms arise as exact solutions to the Hamilton-Jacobi equations for the counterterm action. In other words, such actions arise after resumming an infinite set of terms. We will be content in gaining some insight by studying Einstein gravity leaving a generalization to higher derivative gravity for future work.

\subsection{Counterterm actions in AdS$_4$/CFT$_3$}
 The Gauss-Codazzi decomposition of Einstein equations can be written as \cite{kraus}
\be G_{ab}=G_{ab}(\g)+\hat{n}^{m}\nabla_{m}\Pi_{ab}-\frac{1}{2}\g_{ab}(\frac{\Pi^{2}}{d-1}-\Pi_{cd}\Pi^{cd})+\frac{1}{d-1}\Pi_{ab}\Pi,
\ee
\be
G_{am}\hat{n}^{m}=-\nabla^{b}\Pi_{ab},\ee \be G_{\m\n}\hat{n}^{\m}\hat{n}^{\n}=\frac{1}{2}[\frac{1}{d-1}\Pi^{2}-\Pi_{ab}\Pi^{ab}-\R(\g)],
\ee
where $\hat{n}^{\m}$ is the outward pointing
normal to the boundary.
Considering an AdS bulk, we have \beq G_{ab}=\frac{1}{2}\frac{d(d-1)}{L^{2}}\g_{ab},\nonumber\\G_{am}\hat{n}^{m}=0,\nonumber\\
G_{\m\n}\hat{n}^{\m}\hat{n}^{\n}=\frac{1}{2}\frac{d(d-1)}{L^{2}}\,.\eeq \par
We now proceed to solve the bulk equations perturbatively in the Fefferman-Graham coordinates. The divergent part of the normal derivatives can be expressed in terms of the intrinsic boundary data \cite{bala} which we implement through the Hamiltonian constraint \be
\frac{1}{d-1}\tilde{\Pi}^{2}-\tilde{\Pi}_{ab}\tilde{\Pi}^{ab}=\frac{d(d-1)}{L^{2}}+\R,\ee and insist that the counterterm action must be intrinsic to the boundary as\be
\tilde{\Pi}^{ab}=\frac{2}{\sqrt{|\g|}}\frac{\d}{\d\g_{ab}}\int d^{d}x\sqrt{|\g|} S, \ee
where $S$ is identified as the counterterm Lagrangian, which can fully determine the counterterm in the absence of anomalies and singularities. A perturbative recursion relation solving the Hamiltonian constraint can be set up as shown in appendix A. The counterterm action can be expanded in a derivative expansion. For example the zero, two and four derivative terms are given by \cite{kraus},
 \beq
S_0&=& \frac{d-1}{L} \,,\\
S_1&=& \frac{L}{2(d-2)} \R\,,\\
S_{2} &=& \frac{L^{3}}{2(d-4)(d-2)^{2}} \,[\R_{2}-\frac{d}{4(d-1)}\R^{2}]\,.
\eeq
Note that in $d=2,4$ we have recovered the pole terms which arises due to the conformal anomaly which signal a breakdown in the expansion. In odd dimensions, such pole terms are absent signalling the absence of anomalies. This means that one can in principle resum the entire series of terms arising from solving the Hamiltonian constraint order by order.
Resumming an infinite set of terms using this formalism, however,  seems impossible! One needs to start with a reasonable guess for the solution. In \cite{js} it was shown that if one considers
a DBI type counterterm \footnote{Without the $\R_{ab}$, this is the Mann-Lau counterterm \cite{mann,lau}. This action also coincides with the zero central charge \cite{js} extension of NMG \cite{tekinnmg}. The one parameter ambiguity in \cite{sss} gets fixed to the DBI value when we re-write the leading non-conformal correction in terms of the Cotton tensor.} :
\begin{equation} \label{eqm1}
I_{ct}=-\frac{2L^2}{\ell_P^2} \sqrt{-{\rm det}(\R_{ab}-\frac{1}{2}\R
  \g_{ab}-\frac{1}{L^2} \g_{ab})}\,,
\end{equation}
then this cancels off the cut-off dependence in the Euclidean action on $S^3$ or $S^1\times S^2$ (also for $H^3, S^1 \times H^2, R^3$ ) exactly and as such is valid for any radius. Quite remarkably, it can be shown \cite{mlink} that these are exact solutions to the Hamiltonian constraint provided we consider the boundaries to be $S^3$, $S^1\times S^2$,  $H^3, S^1 \times H^2, R^3$. Incidentally, this Lagrangian coincides with the CDJ Lagrangian \cite{cdj, peldan} in 2+1 dimensions and is the Legendre transform of the Ashtekar Hamiltonian \cite{peldan}. Of course, since we wish to consider fluctuations away from $S^3$ or $S^1 \times S^2$, one needs to consider the systematics of the corrections to the counterterm action displayed in eq.(\ref{eqm1}). Using the results of \cite{kraus} it can be shown that at six derivative order, the counterterm action is given by
 \beq \label{cubic}
S_{3}(\g)=\frac{L^{5}}{(d-6)(d-2)^{4}}\left[\R_{3}-\frac{3d}{4(d-1)}\R\R_{2}+\frac{d^2+4d-4}{16(d-1)^{2}}\R^{3}-\frac{d-2}{d-4}(\frac{1}{2}C^{2}+\R^{ik}\R^{mn}W_{mkin})\right]\nonumber \\
\eeq
where $W_{ijkl}$ is the Weyl tensor, $C_{ijk}$ is the Cotton tensor which
can be expressed as \be C_{ijk}=\nabla_{k}\R_{ij}-\nabla_{j}\R_{ik}+\frac{1}{2(d-1)}(g_{ik}\nabla_{j}\R-g_{ij}\nabla_{k}\R),\ee and `measures' the deviation from conformalilty of the space in 3d and
$C^{2}=C^{ijk}C_{ijk}$.  Setting $d=3$ eq.(\ref{cubic}) coincides with the expansion of the action in eq.(\ref{eqm1}) upto this order with the only additional term being the $C^2$ term (since the Weyl tensor identically vanishes in $d=3$). Thus the exact counterterm action valid for manifolds away from conformality can be presented as an expansion in powers of the Cotton tensor:
\beq \label{eqm2}
I_{ct}&=&-\frac{2L^2}{\ell_P^2} \sqrt{-{\rm det}(\R_{ab}-\frac{1}{2}\R
  \g_{ab}-\frac{1}{L^2} \g_{ab})}+O(C^2)\,,\\
&=&\frac{2}{L \lp^{2}}\int d^{3} x\sqrt{\gamma} (1+\frac{1}{2}L^{2}\mathcal{R}-\frac{1}{2}L^{4}(\mathcal{\R}_{2}-\frac{1}{2}\R^{2})-\frac{1}{24} L^{6}\mathcal{R}^{3}+\frac{1}{4}L^{6}\mathcal{R}\mathcal{R}_{2}-\frac{1}{3}L^6 \mathcal{R}_{3})^{1/2}+O(C^2)\,,\nonumber\\
\eeq
where in the second line we have expanded the determinant and introduced the notation:
$
\R_n={\rm tr}\,(\R^n)\,.
$ Note that the coefficient $2L^2/\lp^2$ is the finite part of the Euclidean partition function on $S^3$. Further, we anticipate the correction at eight-derivative order to be of the form $O(C^2 R)$ but its precise form is not known which will limit the order upto which we will be able to expand our action in the `dilaton' field \footnote{ We have however, explicitly verified that on a conformally flat boundary, the difference between the eight-derivative $O(C^2 R)$ terms and the eight-derivative terms coming from the DBI-counterterm vanishes. A preliminary investigation \cite{ks} suggests that the corrections at eight-derivative order are remarkably simple and of the form $$-\frac{1}{10} (\nabla^m C_{ijm})^2+\frac{2}{5} (G S)^{a b}\nabla^m C_{abm}+\frac{1}{10}C_{ijk}C^{ijk}R-\frac{1}{5}(C^{ijk}\frac{\delta C_{ijk}}{\delta \gamma^{ab}})S^{ab}$$ where $G$ is the Einstein tensor and $S$ is the Schouten tensor. Thus even at this order the correction features the Cotton tensor and would hence vanish for a conformal boundary. We expect this feature to persist at an arbitrary order although we do not know of a proof. Thus although some of the statements we will make in what follows are conservative, we believe that our findings are more generally valid.}.
Now consider the case where the boundary is $S^3$. The AdS$_4$ metric is
\be
ds^{2}=\frac{dr^{2}}{\frac{r^{2}}{L^{2}}+1}+r^{2}d\Omega^{2}\,,\\
\ee
where, $d\Omega^{2}$ is the metric of unit three sphere (the usual boundary in these coordinates is at $r\rightarrow \infty$).
\be
d\Omega^{2}=d\theta^{2}+\sin(\theta)^{2}d\phi^{2}+\sin(\theta)^{2}\sin(\phi)^{2}d\psi^{2}\,.\\
\ee
Since $\R\sim 1/r^2$, if we considered the neighbourhood of $r=0$, it would be the $O(\R^3)$ terms that would dominate inside the square-root in eq.(\ref{eqm2}). But these are precisely the terms that are left invariant under global Weyl transformations! Furthermore, note that in the calculation of the finite part of the Euclidean partition function on $S^3$, it is precisely this part of the geometry that contributes! With these coincidences, we have enough motivation to look at the fluctuations around the $S^3$ metric more closely.
\subsection{ $S^{3}$ boundary}
Motivated by our calculations for even-dimensional CFTs, we promote the cut-off to a field of the form
\be
r= L  e^{\sigma(\theta,\phi,\psi)}\,.
\ee
The total action is then expanded around $r=0$, i.e $\sigma \rightarrow-\infty$ giving\,,
\be
I_{tot}=I_{bulk}+I_{GH}+I_{ct}\,,\\
\ee
where,
\begin{align}
\begin{split}
I_{bulk}&=-\frac{1}{2\lp^{2}}\int  d^{4}x\sqrt{g} (R+\frac{6}{L^2}\,),\quad I_{Gh}=-\frac{1}{\lp^{2}}\int d^{3}x\sqrt{\gamma}\, \mathcal{K} \,,\\
I_{ct}&=\frac{2}{L \lp^{2}}\int d^{3} x\sqrt{\gamma} (1+\frac{1}{2}L^{2}\mathcal{R}-\frac{1}{2}L^{4}(\mathcal{R}_{2}-\frac{1}{2}R^{2})-\frac{1}{24} L^{6}\mathcal{R}^{3}+\frac{1}{4}L^{6}\mathcal{R}\mathcal{R}_{2}-\frac{1}{3}L^6\mathcal{R}_{3})^{1/2}\,.
\end{split}
\end{align}
Here\,, $\mathcal{K}$, $\mathcal{R}_{2}$ and $\mathcal{ R}_{3}$ is defined on the boundary as,
\begin{align}
\begin{split}
\mathcal{R}_{2}&=\mathcal{R}^{a}_{b}\mathcal{R}^{b}_{c}\,,\quad \mathcal{R}_{3}=\mathcal{R}^{a}_{b}\mathcal{R}^{b}_{c}\mathcal{R}^{c}_{a}\,,\quad \mathcal{K}=g^{ab}\nabla_{a} n_{b}\,,
\end{split}
\end{align}
$n_{b}$ unit normal defined on the boundary and $\gamma$ is the boundary metric and $a$,$b$ denote the boundary indices.\par
As observed, only the $\mathcal{R}^{3}$ order terms inside the square-root in the counterterm action contributes to the leading order; other terms are suppressed  by exponential factors and die off in the  $\sigma(\theta,\phi,\psi)\rightarrow -\infty$ limit. We find,
\be
I=\frac{2L^2}{\lp^2}\int d^{3}x\sqrt{\gamma}(-\frac{1}{24} \mathcal{R}^{3}+\frac{1}{4}\mathcal{R}\mathcal{R}_{2}-\frac{1}{3}\mathcal{R}_{3})^{1/2}\,.
\ee
To simplify the calculation we make a conformal transformation of the metric,
\be
\hat{ds^{2}}=\frac{1}{r^{2}}ds^{2}\,,\\
\ee
where, the conformal factor is $ e^{-2\sigma(\theta,\phi,\psi)}$. Then the induced metric is
\be \label{met1}
d\hat{\gamma}^{2}= \partial_{a}\sigma\partial_{b} \sigma dx^{a}dx^{b}+d\hat{\Omega^{2}}+\mathcal{O}(\sigma^{4})\,,\\
\ee
where, we will use hatted symbols for the quantities calculated using this metric and $d\hat\Omega^{2}$ is the metric of unit three sphere. At this stage note that the form of the above metric immediately tells us that the $O(C^2)$ terms in the counterterm action will be at least $O(\sigma^4)$ so that we can trust our analysis from the DBI counterterm at least upto $O(\sigma^3)$. In order to do better than this one would need to consider higher order terms in the Cotton tensor which we will not consider here.

\par
Then $\mathcal{\hat{R}}^{3}$, $\mathcal{\hat{R}}\mathcal{\hat{R}}_{2}$ and $\mathcal{\hat{R}}_{3}$ are evaluated upto $\mathcal{O}(\sigma^{3})$ and then they are conformally transformed back to give the original quantities. We will  quote the result for the Christoffel symbol and the Riemann tensor below.  Bar denotes a quantity calculated using the unit sphere.
\be
\hat \Gamma^{a}_{bc}=  \bar\Gamma^{a}_{bc}+\bar\nabla^{a}\sigma\bar\nabla_{b}\bar\nabla_{c}\sigma+\mathcal{O}(\sigma^{4})\,.\\
\ee
\be
\hat \R_{abc}{}^{d}=\bar \R_{abc}{}^{d}-\bar\nabla^{d}\sigma\bar \R_{abc}{}^{e}\bar\nabla_{e}\sigma+\bar\nabla_{b}\bar\nabla^{d}\sigma\bar\nabla_{a}\bar\nabla_{c}\sigma-
\bar\nabla_{a}\bar\nabla^{d}\sigma\bar\nabla_{b}\bar\nabla_{c}\sigma+\mathcal{O}(\sigma^{4})\,.\
\ee
\be
\hat \R^{b}_{d}=\bar \R^{b}_{d}-\bar \nabla^{b}\sigma \bar \nabla_{d} \sigma -\delta^{b}_{d}(\bar \nabla\sigma)^{2}+\bar \Box \sigma \bar\nabla^{b}\bar\nabla_{d}\sigma-\bar\nabla^{b}\bar\nabla^{c}\sigma\bar\nabla_{c}\bar\nabla_{d}\sigma+\mathcal{O}(\sigma^{4})\,.\\
\ee
\be
\hat \R=6-4 (\bar\nabla\sigma)^{2}+(\bar \Box\sigma)^{2}-(\bar\nabla_{a}\bar\nabla_{b}\sigma)^{2}+\mathcal{O}(\sigma^{4})\,.\\
\ee
For unit $S^3$,
\begin{align}
\bar\R_{abcd}=(\bar\gamma_{ac}\bar\gamma_{bd}-\bar\gamma_{ad}\bar\gamma_{bc})\,, \quad \bar   \R^{b}_{d}=2\delta^{b}_{d}\,.\\
\end{align}
$\bar\gamma_{ac}$ is the unperturbed part of the metric.\par
Now we use the transformation rule for the Ricci tensor under conformal transformations \cite{wald},
\be\label{conft}
\R_{ab}=\hat \R_{ab}-\nabla_{a}\nabla_{b}\sigma-g_{ab}\Box \sigma+\nabla_{a}\sigma\nabla_{b}\sigma-g_{ab}(\nabla\sigma)^{2}\,.\\
\ee
Here, all the un-barred covariant derivatives are taken with respect to the metric in eq.(\ref{met1}). Then we have,
\begin{align}
\begin{split}
\mathcal{R}^{a}_{b}&=\frac{e^{-2\sigma}}{L^{2}}(2\delta^{a}_{b}-2\delta^{a}_{b}(\bar\nabla \sigma)^{2}+\delta^{a}_{b}(\bar\nabla \sigma)^{2}\bar\Box\sigma+\bar\nabla^{a}\bar\nabla_{b}\sigma(\bar\nabla \sigma)^{2}+\bar\nabla^{a}\bar\nabla_{b}\sigma\bar\Box\sigma-\bar\nabla^{a}\bar\nabla_{b}\sigma+\bar\nabla^{a}\sigma\bar\nabla_{c}\sigma\bar\nabla^{c}\bar\nabla_{b}\sigma\\&
+\delta^{a}_{b}\bar\nabla^{c}\sigma\bar\nabla^{d}\sigma\bar\nabla_{d}\bar\nabla_{c}\sigma-\delta^{a}_{b}\bar\Box\sigma-\bar\nabla^{a}\bar\nabla_{c}\sigma\bar\nabla^{c}\bar\nabla_{b}\sigma)+\mathcal{O}(\sigma^{4})\,.\\
\end{split}
\end{align}
\be
\R= \frac{e^{-2\sigma}}{L^{2}}(6-6(\bar\nabla\sigma)^{2}+4(\bar\nabla \sigma)^{2}\bar\Box\sigma-4\bar \Box \sigma+(\bar \Box \sigma)^{2}+\bar \nabla^{a}\sigma \bar\nabla_{c}\sigma\bar\nabla^{c}\bar\nabla_{a}\sigma-(\bar\nabla_{a}\bar\nabla_{b}\sigma)^{2})+\mathcal{O}(\sigma^{4})\,.\\
\ee
We reiterate that all the covariant derivatives denoted with a bar are taken with respect to metric the of a unit $S^3$.
Then upto\,$ \mathcal{O}(\sigma^{4})$,
\be
I=\frac{2 L^{2}}{\lp^{2}}\int d^{3}x \sqrt{\bar\gamma}\,\sqrt{1+\mathcal{L}_{1}+\mathcal{L}_{2}+\mathcal{L}_{3}}\,,\\
\ee
where the suffixes  denote the order of $\sigma$ involved in the corresponding terms.
\begin{align}
\begin{split}
\mathcal{L}_{1}&=-2 \bar\Box \sigma\,,\quad \mathcal{L}_{2}=-2 (\bar\nabla \sigma)^{2}+2(\bar\Box \sigma)^{2}-(\bar\nabla_{a}\bar\nabla_{b}\sigma)^{2}\,,\\
\mathcal{L}_{3}&=4(\bar\nabla \sigma)^{2}\bar\Box \sigma+2 \bar\nabla^{a}\sigma\bar\nabla^{b}\sigma\bar\nabla_{a}\bar\nabla_{b}\sigma-\frac{2}{3}\bar\nabla^{a}\bar\nabla_{b}\sigma\bar\nabla^{b}\bar\nabla_{c}\sigma\bar\nabla^{c}\bar\nabla_{a}\sigma-\frac{4}{3}(\bar\Box\sigma)^{3}+2(\bar\nabla_{a}\bar\nabla_{b}\sigma)^{2}\bar\Box\sigma\,.\\
\end{split}
\end{align}
Integrating by parts we derive (upto surface terms),
\begin{align} \label{deri}
\begin{split}
(\bar\nabla_{a}\bar\nabla_{b}\sigma)^{2}&=(\bar\Box \sigma)^{2}-2 (\bar\nabla \sigma)^{2}\,,\quad\bar\nabla^{a}\sigma\bar\nabla^{b}\sigma\bar\nabla_{a}\bar\nabla_{b}\sigma=-\frac{1}{2}(\bar\nabla\sigma)^{2}\bar\Box\sigma\\  \bar\nabla^{a}\bar\nabla_{b}\sigma\bar\nabla^{b}\bar\nabla_{c}\sigma\bar\nabla^{c}\bar\nabla_{a}\sigma&=\frac{3}{2}(\bar\nabla_{a}\bar\nabla_{b}\sigma)^{2}\bar\Box\sigma-\frac{1}{2}(\bar\Box\sigma)^{3}+\frac{3}{2}(\bar\nabla \sigma)^{2}\bar\Box \sigma\,.\\
\end{split}
\end{align}
If we expand $I$ upto $ \mathcal{O}(\sigma^{3})$ and use the above relations we get
\be
I_{ct}=\frac{2L^2}{\lp^2}\int d^3 x \sqrt{\bar\gamma}(1+\bar\nabla_a J^a)\,,
\ee
where
\be
J^a=-\bar\nabla^a \sigma+\frac{1}{2}(\bar\nabla^a\sigma \bar\nabla^2 \sigma-\bar\nabla^a \bar\nabla^b \sigma \bar\nabla_b \sigma)-\frac{1}{3}[\bar\nabla^c\bar\nabla^a \sigma \bar\nabla_b \sigma \bar\nabla^b \bar\nabla_c\sigma+\bar\nabla^a\sigma (\bar\nabla\bar\nabla\sigma)^2+\bar\nabla^a\sigma(\bar\nabla\sigma)^2]\,.
\ee
In other words upto $O(\sigma^3)$ we get total derivatives.
Was this result expected? We can attempt to draw an analogy with what happens in 3+1 dimensions. There we have a conformal anomaly. There are two pieces, one proportional to the 4d-Euler density which is a topological quantity and the other to the Weyl-square which is invariant under Weyl transformations. The Euler density would be invariant under metric perturbations. The counterterms under consideration in 3-dimensions consist of a DBI type term and a series of corrections whose leading term is the Cotton square. The 3d $C^2$ term (and subsequent higher order terms) plays the role of the 4d Weyl-square while it seems that upto $O(\sigma^3)$ the DBI counterterm plays the role of the Euler density. We should point out immediately, that we do not expect terms at higher order say $O(\sigma^4)$ to be total derivatives since the DBI-type term is not expected to a topological term in 3-dimensions. Nonetheless, it is very curious that this analogy appears to exist (upto an order) with what happens in 4-dimensions. Let us also emphasise that this does not mean that there is an anomaly in 3-dimensions. The reason why we get a non-zero trace in even dimensions from the counterterm viewpoint is that the global Weyl invariant terms come with pole factors. There are no poles in odd dimensions and although we have extracted a piece from the counterterm that is invariant under scalings, this term will not lead to a non-zero trace. Rather since this piece is invariant under scalings, it does appear to point to the universal nature of the F-theorem conjecture.

\subsection*{Weyl transformation of the counterterm}
As shown in appendix B, one can extract the WZ action from the pole terms in even dimensions by doing a Weyl transformation. Thus it is natural to ask what happens to the DBI counterterm under Weyl transformation. Since the $C^2$ terms (which would have contributed to $O(\sigma^4)$ in the previous discussion) will be Weyl invariant we can just focus on the DBI part of the counterterm action.
If we perform a Weyl transformation, eq.(\ref{conft}), of the boundary metric $\gamma_{ab}\rightarrow e^{2\tau}\gamma_{ab}$, then the DBI counterterm considered around $S^3$ in the small $r$ limit changes to
\be
I_{R^3}=\frac{2 L^{2}}{\lp^{2}}\int d^3 x \sqrt{\bar\gamma} (1-\bbox\tau)^{3/2}\left[{\rm det}\left(\delta_{a}^c+\frac{\bar\nabla_a \bar\nabla^c\tau-\bar\nabla_a\tau \bar\nabla^c\tau}{1-\bbox\tau}\right)\right]^{1/2}\,.
\ee
Now,
\be
{\rm{det}}~ M_{3\times 3}=\frac{1}{3}{\rm Tr}(M^{3})-\frac{1}{2}{\rm Tr}(M^{2}){\rm Tr}M+\frac{1}{6}({\rm Tr} (M))^{3}\,,\\
\ee
with
$$M= \delta_{a}^c\, (1-\bbox \tau)+\bar\nabla_a\bar\nabla^c\tau-\bar\nabla_a\tau\bar\nabla^c\tau.$$
\begin{align}
\begin{split}
{\rm Tr}(M)&=3-2\bbox \tau-(\bnabla \tau)^{2}\,,\\
{\rm Tr}(M^{2})&=3-4\bbox\tau+(\bbox\tau)^{2}-2(\bnabla\tau)^{2}+(\bnabla_{a}\bnabla_{b}\tau)^{2}\\
&-2\bnabla^{a}\tau\bnabla^{b}\tau\bnabla_{a}\bnabla_{b}\tau +2(\bnabla\tau)^{2}\bbox\tau+(\bnabla\tau)^{4}\,,\\
{\rm Tr}(M^{3})&=3-6\bbox\tau+3[(\bbox\tau)^{2}-(\bnabla\tau)^{2}+(\bnabla_{a}\bnabla_{b}\tau)^{2}]\\
&+(\bnabla_{a}\bnabla_{b}\tau)^{3}-6\bnabla^{a}\tau\bnabla^{b}\tau\bnabla_{a}\bnabla_{b}\tau-3\bbox\tau(\bnabla_{a}\bnabla_{b}\tau)^{2}+6\bbox\tau(\bnabla\tau)^{2}\\
&-3[\bnabla_{a}\bnabla_{b}\tau\bnabla^{c}\tau\bnabla_{c}\bnabla^{a}\tau\bnabla^{b}\tau-2\bbox\tau\bnabla^{a}\tau\bnabla^{b}\tau\bnabla_{a}\bnabla_{b}\tau+(\bbox\tau)^{2}(\bnabla\tau)^{2}-(\bnabla\tau)^{4}]\\
&+3[(\bnabla\tau)^2 \bnabla_a\bnabla_b \tau\bnabla^a\tau\bnabla^b\tau-\bbox\tau(\bnabla\tau)^{4}]-(\bnabla\tau)^6\,.
\end{split}
\end{align}
Now using the expressions above we get
\begin{align}
\begin{split}
{\rm det}~M&=1-2\bbox\tau+\frac{1}{2}[3(\bbox\tau)^2-2(\bnabla\tau)^2-(\bnabla_a\bnabla_b\tau)^2]\\
&+\frac{1}{3}[(\bnabla_a\bnabla_b\tau)^3-(\bbox\tau)^3+3 \bnabla_a\bnabla_b\tau \bnabla^a\tau\bnabla^b\tau+3\bbox\tau(\bnabla \tau)^2]\\
&+\frac{1}{2}[(\bnabla\tau)^2(\bnabla_a\bnabla_b\tau)^2-(\bnabla\tau)^2(\bbox\tau)^2-2\bnabla _a\bnabla _b\tau\bnabla ^c\tau\bnabla _c\bnabla ^a\tau\bnabla ^b\tau]\,.
\end{split}
\end{align}
Note that remarkably, all the $O(\tau^5)$ and $O(\tau^6)$ terms have canceled! This is an exact result.

If we expanded the square root of this term to $O(\tau^4)$ we get
\be\sqrt{{\rm det}~M}=1+\L_{1}+\L_{2}+\L_{3}+\L_{4}+O(\tau^5)\,,\ee where
\begin{align}
\begin{split}
\L_{1}&=-\bbox\tau,\quad  \mathcal{L}_{2}=-\frac{1}{2} (\bnabla \tau)^{2}+\frac{1}{4}(\bbox \tau)^{2}-\frac{1}{4}(\bnabla_{a}\bar\nabla_{b}\tau)^{2}\,\\
\mathcal{L}_{3}&=\frac{1}{2} \bar\nabla^{a}\tau\bar\nabla^{b}\tau\bar\nabla_{a}\bar\nabla_{b}\tau+\frac{1}{6}(\bar\nabla_{a}\bar\nabla_{b}\tau)^{3}+\frac{1}{12}(\bar\Box\tau)^{3}-\frac{1}{4}(\bar\nabla_{a}\bar\nabla_{b}\tau)^{2}\bar\Box\tau\,,\\
\L_{4}&=\frac{5}{96}(\bbox\tau)^{4}-\frac{1}{8}(\bbox\tau)^{2}(\bnabla\tau)^{2}-\frac{1}{8}(\bnabla\tau)^{4}+\frac{1}{2}\bbox\tau(\bar\nabla^{a}\tau\bar\nabla^{b}\tau\bar\nabla_{a}\bar\nabla_{b}\tau)-\frac{3}{16}(\bbox\tau)^{2}(\bnabla_{a}\bnabla_{b}\tau)^{2}\\&+\frac{1}{8}(\bnabla\tau)^{2}(\bnabla_{a}\bnabla_{b}\tau)^{2}-\frac{1}{32}[(\bnabla_{a}\bnabla_{b}\tau)^{2}]^2-\frac{1}{2}(\bnabla _a\bnabla _b\tau\bnabla ^c\tau\bnabla _c\bnabla ^a\tau\bnabla ^b\tau)+\frac{1}{6}\bbox\tau(\bnabla_{a}\bnabla_{b}\tau)^{3}\,.\\
\end{split}
\end{align}
Using equation (\ref{deri})  we get total derivatives from, $\L_{1}$,\,$\L_{2}$,\,$\L_{3}$. For $\L_{4}$ we use the following relation obtained by integrating by parts.
\begin{align}
\begin{split}
\bnabla _a\bnabla _b\tau\bnabla ^c\tau\bnabla _c\bnabla ^a\tau\bnabla ^b\tau&= \frac{1}{2}(\bbox\tau)^{2}(\bnabla\tau)^{2}+\bbox\tau(\bnabla ^a\tau\bnabla ^c\tau\bnabla _a\bnabla _c\tau)-(\bnabla \tau)^4-\frac{1}{2}(\bnabla\tau)^{2}(\bnabla_{a}\bnabla_{b}\tau)^{2}\,.\\
\end{split}
\end{align}
Finally we obtain ,
 \begin{align}
\begin{split}\label{l4}
 \L_{4}&=\frac{5}{96}(\bbox\tau)^{4}-\frac{3}{8}(\bbox\tau)^{2}(\bnabla\tau)^{2}+\frac{3}{8}(\bnabla\tau)^{4}-\frac{3}{16}(\bbox\tau)^{2}(\bnabla_{a}\bnabla_{b}\tau)^{2}\\&+\frac{5}{8}(\bnabla\tau)^{2}(\bnabla_{a}\bnabla_{b}\tau)^{2}-\frac{1}{32}[(\bnabla_{a}\bnabla_{b}\tau)^{2}]^2+\frac{1}{6}\bbox\tau(\bnabla_{a}\bnabla_{b}\tau)^{3}\,.\\
\end{split}
\end{align}
So the action becomes,
\be\label{imp}
I_{R^{3}}=\frac{2 L^{2}}{\lp^{2}}\int d^{3}x \,\sqrt{\bar\gamma}\L_{4}\,.\\
\ee
Let us again point out that the coefficient outside this action is the finite part of the Euclidean partition function on $S^3$. Furthermore, there are no scales in $\L_4$ as it originates from an action that is manifestly invariant under scale transformations. 
It will be interesting to see if we can use field theoretic arguments from an action like this to prove the F-theorem along the lines of \cite{Komargodski:2011vj}. Note for instance that if we did a naive derivative expansion of the above action then the leading order contribution would be proportional to $(\bar\nabla \tau)^4$. However, derivatives involving $\bar\nabla$'s above are dimensionless made from the metric of the unit $S^3$. This is because we have absorbed the length dimension of the radial cutoff by the AdS curvature. Since the radial cutoff is interpreted as some
Wilsonian renormalization scale $\mu_R$, this suggests that a derivative expansion of the counterterm action is in fact an expansion in $\mu_R/R$,
where $R$ is the radius of the sphere the CFT is placed on. %JH%

We could have also performed a conformal transformation from $S^3$ to $R^3$ of the form $$ ds_{S^3}^2=\frac{4}{(1+x_i x_i)^2} dx_i dx_i$$ and redefined $e^{2\hat\tau}=e^{2\tau}\frac{4}{(1+x_i x_i)^2}$. In terms of $\hat\tau$, the action takes the form:
\be
\frac{2 L^{2}}{\lp^{2}}\int d^3 x\, \sqrt{\frac{1}{3}(t_2-t_4)+\frac{1}{2}(t_6-t_5-2 t_3)}\,,
\ee
where the $t_i$'s are defined through eq.(\ref{t6}) with $\sigma$ replaced with $\hat\tau$. Furthermore, the integrand is of the form $\sqrt{O(\partial^6)}$ so that clearly there is no scale even if we worked with dimensionful coordinates. Also, this calculation results in a manifestly nonlocal action as there is no possibility of a local expansion in field order in terms of $\hat\tau$.

%However, the meaning of derivative expansion in this case  is obscure to us since 
%$\bar\nabla$'s above are dimensionless made from the metric of the unit $S^3$.

%\begin{align}
%\begin{split}
%J^a&=\frac{2L^{2}}{\lp^{2}}\int d^{3}x \sqrt{\bar\gamma}(1-\bar\Box\tau+\frac{1}{2}(\bar\nabla_{b}(\bar\nabla^{b}\sigma\bar\Box\sigma)-\bar\nabla_{a}(\bar\nabla_{b}\sigma\bar\nabla^{a}\bar\nabla^{b}\sigma))\\&-\frac{1}{3}(\bar\nabla_{c}(\bar\nabla^{a}\bar\nabla^{c}\sigma\bar\nabla_{b}\sigma\bar\nabla^{b}\bar\nabla_{a}\sigma)+\bar\nabla_{a}(\bar\nabla^{a}\sigma(\bar\nabla_{b}\bar\nabla_{c}\sigma)^{2}))+\bar \nabla_{a}(\bar\nabla^{a}\sigma(\bar\nabla^{b}\sigma)^{2})\,.\\
%\end{split}
%\end{align}

%\subsection{$H^3$ boundary}

%\subsection{$R\times H^2$ boundary}

\section{Discussion}
In this paper we studied the AdS/CFT origin of dilaton actions arising due to broken conformal symmetry that have
played a key role in the understanding of $c$-theorems in 1+1 and 3+1 dimensions in recent times. We showed that identifying the radial coordinate of
the cut-off surface in AdS in a certain way with the dilaton, or in other words promoting the cut-off to a field, the derivative expansion of this field coincides with the dilaton action in flat space.
In particular, we recover the WZ action which is associated to a conformal anomaly that is in complete agreement with the holographic
anomaly obtained in the conventional way \cite{Henningson:1998gx} in the $N\rightarrow \infty$ limit. We explicitly showed that in Gauss-Bonnet theory, where one can distinguish the Euler anomaly coefficient from the other anomaly coefficients associated with Weyl invariants, the coefficient of the dilaton action was proportional to the Euler anomaly. Furthermore, the emergence of the dilaton action from our procedure using counterterms persists even for the new massive gravity theory whose equations of motion are not two derivative, unlike the Gauss-Bonnet theory.

In a way, one might consider our results as confirmation of the consistency of the AdS/CFT correspondence,
since it is expected that the WZ action should be naturally reproduced in the
CFT path integral when one performs a Weyl rescaling
$\eta \to e^{2\sigma}\eta$ in even boundary CFT dimensions. This fact about holographic actions
has been pointed out
before in the
literature, and demonstrated explicitly in Einstein gravity in four dimensions
using diffeomorphism between different Fefferman-Graham coordinates \cite{Imbimbo:1999bj,Manvelyan:2001pv}.
However, our approach differs in one crucial aspect: we focus our attention entirely on the cut-off surface,
and that has led to some new insights. One observation is
the surprising fact that the counterterms responsible for canceling power law divergences,
in fact, carries full knowledge of the anomaly, which is associated with a
logarithmic divergence. More importantly, given that the counterterms are supposedly
solutions of the Hamilton-Jacobi (HJ) equation \cite{de Boer:1999xf} that guarantees
diffeomorphism invariance along the radial direction which in turn, has been intimately related
to RG Polchinski equation \cite{de Boer:1999xf, Heemskerk:2010hk, Faulkner:2010jy},
our approach naturally opens up the way
to understand $W[\sigma]$ at finite RG scale in a generic holographic RG flow. All that is
needed is to place a boundary coordinate dependent cut-off surface
in the interior of AdS. Of course, to achieve that, it is not adequate
to take only the first few terms in the derivative expansion of
these counterterms as in the case when the cut-off is sufficiently
close to the AdS boundary. Instead, we need a full
non-linear solution of the HJ equation.
This approach is exactly a holographic realization of a space-time dependent
RG scale considered for example in \cite{Komargodski:2011xv} and more recently in
\cite{Lee:2012xb}.

In a generic holographic RG flow, dimensionful
couplings would be turned on. As a result, new counterterms would be needed
for canceling new divergences near the boundary \cite{de Haro:2000xn},
which, when evaluated on a  coordinate dependent cutoff surface,
would in turn give rise to new terms in
the dilaton effective action beyond the WZ action.
These dimensionful couplings are the ingredients needed to produce terms
in the dilaton action with different number of derivatives other than the $d-$derivative terms, and thus
give rise to an overall effective dilaton action that takes exactly the structure
as discussed in \cite{st, Komargodski:2011vj}.
On the other hand, in a holographic RG flow, the bulk geometry has two AdS throats with
different radii connected by some interpolating geometry.
These throats correspond to the UV and IR fixed points respectively.
Our discussion so far focus only on one of the throats. When the contribution from the other throat is taken into account, the overall coefficient becomes the difference between the two Euler anomalies--this fact has also been emphasised recently in \cite{hoyos}. Field theoretically, that corresponds to the discussion in \cite{Luty:2012ww}
where one could subtract off the contribution of the UV WZ terms, so that the dilaton effective
action is sensitive only to the \emph{change} of the central charges along the flow.

Motivated by these observations in even dimensions, we turned to the odd dimensional case and focused on $d=3$. We found that there exists a DBI type counterterm \cite{js,sss} that is an exact solution to the Hamiltonian constraint for $S^3, S^2 \times S^1$ boundaries. When one considers the holographic origin of the finite part of the Euclidean partition function on $S^3$, then one needs to consider the centre of the AdS space. We found that precisely in this limit, the counterterm becomes invariant under global Weyl transformation indicating the fact that the F-theorem proposal is universal. %JH
We note that this property of scale invariance dominated by $\sqrt{h\mathcal{R}^d}$ 
when expanded in the interior of the AdS space
is a general property of the re-summed counter terms in arbitrary odd dimensions, 
at least in the case of Einstein gravity as demonstrated in \cite{sss}.
%JH 
We studied several properties of the counterterm under fluctuations and Weyl transformation. We leave it for future work to see if our results can be used to prove the F-theorem along the lines of \cite{Komargodski:2011vj}. Even if this was achievable in holography, the question would still remain as to how one could extend this in the language of field theory. May be the methods based on symmetry for e.g., in \cite{kurt} will be handy here. One observation that we would like to make here is that in type IIA string theory, if we considered an Euclidean D2-brane, we could wrap it on the $S^3$--this is an instanton. The actions for these objects come with $\exp(-V_3/\ell_s^3 g_s)$ so that in the small $g_s$ limit, they would be suppressed. However, in AdS$_4$, the $V_3$ factor is proportional to $r^3$ so that in the small $r$ limit, the contribution from these instantonic branes would not be suppressed. It could be that our analysis is providing us the action for the scalar moduli for the radial position of the brane. In this sense, we have the action for the Goldstone mode for spontaneously broken conformal symmetry. This interpretation may be helpful in trying to see the field theoretic origin of eq.(\ref{imp}).  It will also be interesting to examine how to use the entanglement entropy proposal of \cite{myersme} and proved for the $d=3$ case in \cite{Casini:2012ei} to shed light on a proof similar to \cite{Komargodski:2011vj}. We hope to return to these
important questions in a future publication.

%Moreover, if we performed the same calculations in this paper for odd dimensional CFTs, it would turn out that expanding the dilaton action close to the boundary would lead to terms that are suppressed in the large cut-off limit.

\section*{Acknowledgments}
We thank Justin David,  R.~Loganayagam, Gautam Mandal, Rob Myers, Ashoke Sen, Nemani Suryanarayana and Tadashi Takayanagi for useful discussions. We are especially grateful to Adam Schwimmer and Zohar Komargodski for discussions and correspondence.
AS is supported in part by a Ramanujan fellowship from Govt. of India.

\appendix

\appendix
\section{ Counterterms from the Hamiltonian constraint}

Here we review\cite{Freidel} in order to set up a recursive procedure to determine the counterterm action--the results will be identical to those obtained in \cite{kraus}. We begin with
$
I_{EH}=I_{bulk}+I_{GH}
$
and define the momentum operators as
\be
\hat{\Pi}_{ab}^{x}=\frac{2}{\sqrt{\g}}\frac{\d}{\d\g^{ab}(x)}\,.
\ee
Thus the operator representation of the Hamiltonian constraint  becomes
\be
\frac{\hat{\H}}{\sqrt{\g}}=-\lp^{2(d-1)}(\hat{\Pi}^{ab}_{x}\hat{\Pi}^{x}_{ab}-\frac{\hat{\Pi}^{2}}{d-1})+\R(\g)+\frac{d(d-1)}{L^{2}}=0\,.
\ee
The momentum constraint becomes
\be \hat{\H}^{b}_{x}=\nabla_{a}\hat{\Pi}^{ab}_{x}=0\,,\ee
which is automatically satisfied by an action that is diffeomorphism invariant. \par
We define
\be \hat{\Pi}^{a}_{b}=\g^{ac}\hat{\Pi}_{bc};\ \ \hat{\Pi}=\hat{\Pi}^{c}_{c}; \ \
\hat{K}_{ab}=\hat{\Pi}_{ab}-\g_{ab}\frac{\hat{\Pi}}{d-1};\ \ \hat{K}^{a}_{b}=\hat{\Pi}^{a}_{b}-\d^{a}_{b}\frac{\hat{\Pi}}{d-1}\,.\ee
We also define $\circ$ as
\be
(\hat{K}\circ\hat{K})_{\m\n}=\g^{\a\b}(\hat{K}_{\m\a}\hat{K}_{\n\b}-\hat{K}_{\m\n}\hat{K}_{\a\b}),\ee and
\be(\hat{K}\circ\hat{K})=\g^{\m\n}(\hat{K}\circ\hat{K})_{\m\n}=(\hat{K}^{\m}_{\a}\hat{K}^{\a}_{\m}-\hat{K}^{2})\,.
\ee
If we
take the quantum mechanical version of the classical HJ equation then the $\hat{\H}$ is such that acting on a state $\Phi=\exp{i\tilde{S}(\g)}$ where $\tilde{S}(\g)$ is the functional of the
boundary metric $\g$ also called the Wheeler de-Witt state, will produce \be \hat{\H}\Phi=0\,.\ee
Putting in the solution of the radial WdW equation into the HJ equation and neglecting the
quantum corrections proportional to $\lp$ we have the operator version of the classical HJ equation as
\be \hat{K}(\tilde{S})\circ\hat{K}(\tilde{S})+\R(\g)+\frac{d(d-1)}{L^{2}}=0.\ee

We now look for the expansion of $\tilde{S}(\g)$ in terms of the functional of the boundary metric having fixed conformal dimension.\textsl{i.e.},\be
\tilde{S}(\g)=\sum\limits_{n=0}^{\infty}S_{n}(\g)\ee
and
\be S_{n}(\rho^{-2}\g)=\rho^{-d+2n}S_{n}(\g)\,.\ee
We expand about $\r=0$ i.e. around the usual boundary, and start at the zeroth order by selecting
\be S_{0}(\g)=c\int\sqrt{\g}\ee
and inserting
into the HJ equation we have 
\be \hat{K}^{b}_{a}\int\sqrt{\g}=\g^{bc}(\hat{\Pi}_{ac}-\g_{ac}\frac{\hat{\Pi}}{d-1})\int\sqrt{\g}.\ee
Now we have 
\be \hat{\Pi}_{ac}\int\sqrt{\g}=-\g_{ac}; \ \ {\rm and} \ \ \hat{\Pi}\int\sqrt{\g}=-d.\ee
Putting all these together we have \be \hat{K}^{a}_{b}\int\sqrt{\g}=\frac{\d^{a}_{b}}{d-1}\ee and \be
\hat{K}(S_{0}(\g))\circ\hat{K}(S_{0}(\g))+\R(\g)+\frac{d(d-1)}{L^{2}}=0\,,\ee
which when expanded about $\rho=0$ gives \be c=\pm\frac{d-1}{L}\,.\ee
Thus\be
S_{0}(\g)=\frac{d-1}{L}\int_{\Sigma}\sqrt{\g}\ee the positive sign of $c$ is chosen to match the classical analysis and we denote the boundary by $\Sigma$. At the next level we put
\be\tilde{S}(\g)=S_{0}(\g)+S(\g)\ee in the HJ equation to obtain \be 2[\hat{K}(S_{0})\circ\hat{K}(S)]=-[\hat{K}(S)\circ\hat{K}(S)]-\R(\g)\,.\ee Now
\be
\hat{K}(S_{0})\circ\hat{K}(S)=-\frac{1}{L}\hat{\Pi}(S)\,.\ee
Thus we obtain
\be \frac{2}{L}\d^{D}_{x}(S)=(\hat{K}(S)\circ\hat{K}(S))+\R(\g)\,,\ee where $\d^{D}_{x}$ is the conformal dimension
operator acting on $S(\g)$. Now the action of $\d^{D}_{x}$ on S may be written as \be \d^{D}_{x}(S)=\frac{1}{\sqrt{\g}}\frac{\d}{\d\phi}S(e^{2\phi}\g)\,,\ee where $\phi$ is the conformal
parameter. We have \be e^{2\phi}=\rho^{-2}\ {\rm and}\ \tilde{S}(\g)=\sum\limits_{n=1}^{\infty}S_{n}(\g)\,.\ee
Thus \be \tilde{S}(\rho^{-2}\g)=\sum\limits_{n=1}^{\infty}\rho^{-d+2n}S_{n}(\g)\,.\ee
Now
\be e^{\phi}=e^{r/L}=1/\r\ee which implies that \be \frac{d\r}{dr}=-\frac{\rho}{L},\ee therefore\be \frac{2}{L}\d^{D}_{x}(S)=2\frac{\partial S}{\partial r}=-2\frac{\r}{L}\frac{\partial
S}{\partial\r}=-2\frac{\r}{L}\partial_{\r}\sum\limits_{n=1}^{\infty}\r^{-d+2n}S_{n}(\g)=2\sum\limits_{n=1}^{\infty}\frac{d-2n}{L}\r^{-d+2n}S_{n}(\g).\ee
On the other hand $\sqrt{\g}R$ transforms
under conformal rescaling
%as \be \g\rightarrow(e^{2r/L})^{d}\g=\r^{-2d}\g,\ee Therefore \be \sqrt{\g}=\r^{-d}\sqrt{\g}.\ee Also \be \R{\g}\rightarrow \r^{2}\R(\g),.\ee
%Thus
\be
\sqrt{\g}\R(\g)\rightarrow \r^{-(d-2)}\sqrt{\g}\R(\g)\,.\ee
Also $ \hat{K}(\tilde{S})\circ\hat{K}(\tilde{S})$ under such rescaling becomes \be
\hat{K}(\tilde{S})\circ\hat{K}(\tilde{S})=\sum\limits_{\a=1}^{\infty}\sum\limits_{\b=1}^{\infty}\int_{\Sigma}\r^{-d+2\a+2\b}\sqrt{\g}[\hat{K}(S_{\a})\circ\hat{K}(S_{\b})].\ee
Thus we obtain recursively \be
\sum\limits_{n=1}^{\infty}2\frac{d-2n}{L}\r^{-d+2n}S_{n}(\g)=\int_{\Sigma}\r^{-d+2}\sqrt{\g}R(\g)+\sum\limits_{\a,\b=1}^{\infty}\int_{\Sigma}\r^{-d+2\a+2\b}\sqrt{\g}[\hat{K}(S_{\a})\circ\hat{K}(S_{\b})]\,.\ee
Comparing the coefficients for all powers of $\r$ for the above identity we have \be 2\frac{d-2}{L}\r^{-d+2}S_{1}(\g)=\r^{-d+2}\int_{\Sigma}\sqrt{\g}\R(\g)\ee implies\be
S_{1}(\g)=\frac{L}{2(d-2)}\int_{\Sigma}\sqrt{\g}\R(\g),\ee and from $2\a+2\b=2n$ we have $\b=n-\a$. Thus \be
\sum\limits_{n=2}^{\infty}\frac{2(d-2n)}{L}\r^{-d+2n}S_{n}(\g)=\sum\limits_{n=2}^{\infty}\r^{-d+2n}\sum\limits_{\a=1}^{n-1}\int_{\Sigma}\sqrt{\g}[\hat{K}(S_{\a})\circ\hat{K}(S_{n-\a})],\ee which reduces
to\begin{align}\boxed{ 2(d-2n)S_{n}(\g)=\sum\limits_{\a=1}^{n-1}\int_{\Sigma}L\sqrt{\g}[\hat{K}(S_{\a})\circ\hat{K}(S_{n-\a})]\,.}\end{align} For example \be
S_{2}(\g)=\frac{1}{2(d-4)}\int_{\Sigma}L\sqrt{\g}[\hat{K}(S_{1})\circ\hat{K}(S_{1})],\ee \be S_{3}(\g)=\frac{1}{d-6}\int_{\Sigma}L\sqrt{\g}[\hat{K}(S_{1})\circ\hat{K}(S_{2})],\ee and \be
S_{4}(\g)=\frac{1}{2(d-8)}\int_{\Sigma}L\sqrt{\g}[2\hat{K}(S_{1})\circ\hat{K}(S_{3})+\hat{K}(S_{2})\circ\hat{K}(S_{2})],\ee \textsl{etc}. One can easily verify that (upto total derivatives) the above recursion relations reproduce the results of \cite{kraus} One obtains at six derivative order\beq
\Pi_{ab}[S_{2}(\g)]\Pi^{ab}[S_{1}(\g)]=\frac{L^{4}}{(d-4)(d-2)^{3}}[2\R_{3}+\frac{d^{2}-9d+6}{4(d-1)}\R\R_{2}+\frac{d(d+2)}{16(d-1)}\R^{3}]\nonumber\\+\frac{L^{4}}{(d-4)(d-2)^{3}}[-2\R^{c}_{a}\nabla_{b}\nabla_{c}G^{ab}-\frac{d}{4(d-1)}\R\Box
\R+\R^{ab}\Box\R_{ab}],\eeq which can be put into the form using the definition of the Cotton tensor $C_{abc}$ as\beq
S_{3}(\g)=\frac{L^{5}\sqrt{\g}}{(d-6)(d-4)(d-2)^{3}}[\frac{d-4}{d-2}\R_{3}-\frac{3d(d-4)}{4(d-1)(d-2)}\R\R_{2}+\frac{d^{3}-20d+16}{16(d-1)^{2}(d-2)}\R^{3}\nonumber\\-\frac{1}{2}C^{2}-\R^{ik}\R^{mn}W_{mikn}].\eeq
Using the above form for the counterterm in $d=3$ we observe that it coincides with the expansion of the DBI counterterm\footnote{Similarly for the $d=5$ case we have from
the above counterterm $\b_{7}=-7/5184$ which coincides with the coefficients that we derive from the simple c-theorem constraint \cite{sss}.}. The form of counterterm can be obtained by noting the variation of the Ricci tensor which is given as \be \frac{\d
\R_{cd}}{\d\g^{ab}}=\frac{1}{2}(\nabla^{l}\nabla_{c}\frac{\d\g_{ld}}{\d\g^{ab}}+\nabla^{l}\nabla_{d}\frac{\d\g_{lc}}{\d\g^{ab}}-\g^{lr}\nabla_{c}\nabla_{d}\frac{\d\g_{lr}}{\d\g^{ab}}-\Box\frac{\d\g_{cd}}{\d\g^{ab}}),\ee
and \be\frac{\d\R}{\d\g^{ab}}=\R_{ab}+(\nabla^{l}\nabla^{d}\frac{\d\g_{dl}}{\d\g^{ab}}-\g^{cd}\Box\frac{\d\g_{cd}}{\d\g^{ab}}),\ee and \be \frac{\d
\R_{2}}{\d\g^{ab}}=2\R^{d}_{a}\R_{db}+2\R^{cd}\frac{\d\R_{cd}}{\d\g^{ab}}.\ee Using these we have \be \Pi_{ab}[S_{1}(\g)]=\frac{L}{d-2}G_{ab},\ee where $G^{ab}$ is the Einstein tensor and \beq
\Pi_{ab}[S_{2}(\g)]=\frac{L^{3}}{(d-4)(d-2)^2}[(2\R^{d}_{a}\R_{db}-\frac{d}{2(d-1)}\R\R_{ab})-\frac{\g_{ab}}{2}(\R_{2}-\frac{d}{4(d-1)}\R^{2})\nonumber\\+\frac{d}{2(d-1)}(\nabla_{a}\nabla_{b}\R-\g_{ab}\Box
\R)-(2\nabla_{c}\nabla_{(b}\R^{c}_{a)}-\frac{\g_{ab}}{2}\Box\R-\Box\R_{ab})].\eeq

\section{Weyl transformation of pole terms}
We consider the pole terms in $d=4$.
\begin{align}
\begin{split}\label{eqn}
I_{pole}=\frac{L^3}{2(d-4)(d-2)^{2}\lp^{3}}\int d^{4}x\sqrt{\g}\,[\R^{ab}\R_{ab}-\frac{d}{4(d-1)}\R^2]\,.
\end{split}
\end{align}
 Using the Weyl transformation of $\R_{ab}$  we get,
\begin{align}
\begin{split}
\hat{\R}_{ab}\hat{\R}^{ab}&=e^{-4\tau}[ \R^{ab}\R_{ab}+(d-2)^{2}t^{b}_{a}t^{a}_{b}+d\,s^{2}-2(d-2)\R_{ab}t^{ab}-2\R s+2(d-2)ts]\,,\\
\hat{\R}^{2}&= e^{-4\tau}[\R^{2}+4(d-1)^{2}(\Box\tau)^{2}+(d-1)^2 (d-2)^{2}(\nabla\tau)^{4}-4\R(d-1)\Box\tau\\&-2\R(d-2)(d-1)(\nabla\tau)^{2}+4(d-1)^{2}(d-2)\Box\tau(\nabla\tau)^{2}]\,,
\end{split}
\end{align}
where,
\begin{align}
\begin{split}
t^{b}_{a}=\nabla^{b}\nabla_{a}\tau-\nabla^{b}\tau\nabla_{a}\tau;\quad
t=\Box\tau-(\nabla\tau)^{2};\quad
s=\Box\tau+(d-2)(\nabla\tau)^{2}\,.
\end{split}
\end{align}
Conformally transforming eq.(\ref{eqn}), and expanding $e^{(d-4)\tau}=1+(d-4)\tau $ we get,
\begin{align}
\begin{split}
I_{pole}&=\frac{L^3}{2(d-4)(d-2)^{2}\lp^{3}}\int d^{4}x\sqrt{\g}\,(1+(d-4)\tau)[\hat{\R}^{ab}\hat{\R}_{ab}-\frac{d}{4(d-1)}\hat{\R}^2]\\
&= \frac{L^{3}}{2(d-2)^{2}\lp^{3}}[\frac{1}{d-4}\int L_{pole1}+ \int L_{pole2}]\,.
\end{split}
\end{align}
Taking into account the non-divergent terms at $d=4$ we have
\be
L_{pole1}=-(d-4)[(d-2)^{2}\Box\tau(\nabla\tau)^{2}+\frac{1}{4}(d-2)^{2}(d-1)(\nabla\tau)^{4}+(d-2)G^{ab}\nabla_{a}\tau\nabla_{b}\tau]\ee
\be L_{pole2}=6\Box\tau(\nabla\tau)^{2}+4(\nabla\tau)^{4}+2(d-2)G^{ab}\nabla_{a}\tau\nabla_{b}\tau\ee
where we have used the following relations obtained by integrating by parts,
\begin{align}
\begin{split}
(\nabla_{a}\nabla_{b}\tau)^{2}&=(\Box\tau)^{2}-R^{ab}\nabla_{a}\tau\nabla_{b}\tau; \quad
\nabla_{a}\tau\nabla_{b}\tau\nabla^{a}\nabla^{b}\tau=-\frac{1}{2}\Box\tau(\nabla\tau)^{2};\quad\\
\Box\nabla_{b}\tau&=\nabla_{b}\Box\tau+\R_{ab}\nabla^{a}\tau\,.
\end{split}
\end{align}
Combining everything we get,
\begin{align}
\begin{split}
I_{pole}=\frac{L^{3}}{8\lp^{3}}\int d^{4}x \sqrt{\g}\left[\tau(\R_{2}-\frac{1}{3}\R^{2})+2G^{ab}\nabla_{a}\tau\nabla_{b}\tau+2(\nabla\tau)^{2}\Box\tau+(\nabla\tau)^{4}\right]\,.
\end{split}
\end{align}
If we compare with \cite{Komargodski:2011vj} then (after setting $\tau_{here}=-\tau_{there}$),  we can identify $I_{pole}=-S_{anomaly}$.

This in fact is a very general way of generating WZ actions. The WZ action $S_{WZ}( X(R(g)),\tau)$, where $X(R)$ is
some non-Abelian anomaly term which are $d$-derivative scalars formed from products of the Riemann curvatures and their derivatives,
is such that
upon an infinitesimal Weyl transformation, $g\to e^{2\sigma}g,\tau\to \tau + \sigma$
\be \label{WZtrans}
\delta_\sigma S_{WZ} = \sigma \sqrt{g} X(R(g)).
\ee
This suggests that one can always generate the WZ action by looking for another scalar function
$f(R(g))$ such that $\delta_\sigma f(R(g)) = \sigma X(R(g))$, then the WZ action
is given by
\be
S_{WZ}(X(R(g)),\tau) = f(R(g)) - f(R(e^{-2\tau}g)) .
\ee
One can easily see that $S_{WZ}$ defined above transforms precisely as given
in (\ref{WZtrans}).

What is special about the Euler anomaly is that if we analytically continue
the number of dimensions to a continuous variable, we have precisely
\be
\delta_\sigma \sqrt{g}(E_4) = (d-4)\sigma \sqrt{g} E_4,
\ee
up to total derivative terms that vanish inside the integral over all space.
Therefore applying the same logic as before, formally $E_4/(d-4)$ plays the role of
the function $f$ above, and the WZ action for $E_4$ should be
\be
S_{WZ}(E_4(g),\tau) = \frac{1}{d-4} \left(E_4(g) - E_4(e^{-2\tau}g)\right),
\ee
which in the $d\to 4$ limit, reduces precisely to the WZ action of the Euler term
as we know it.
In the above, instead of directly working with the Euler anomaly $E_4$,
we studied the combination obtained from the counter terms $R_{ij}^2 - d/(4(d-1))R^2$.
This combination is proportional to $E_4 - W_{ijkl}^2$, with the definition of the
Weyl tensor defined in general dimension. Since $W_{ijkl}$ is conformally
covariant in arbitrary dimensions, the Weyl term hidden inside $f$ is
canceled out when we take $f(R(g)) - f(R(e^{-2\tau}g))$. Therefore, taking the linear
combination above leads to precisely
the same result as taking $f(R)= 1/(d-4) E_4$ alone.


\begin{thebibliography}{99}
\bibitem{mlink} Mathematica notebooks proving this can be made available on request.
\end{thebibliography}
\end{document}